\documentstyle[twoside,fleqn,espcrc2]{article}
\title{Statistical Mechanics and Black Hole Thermodynamics}

\author{S. Carlip\address{Physics Department, University of California
        at Davis \\ Davis, CA 95616, USA}%
        \thanks{Supported in part by National Science Foundation grant
        PHY-93-57203 and U.S.\ Department of Energy grant 
        DE-FG03-91ER40674.}
        }
       
\begin{document}

\begin{abstract}
Black holes are thermodynamic objects, but despite recent progress, 
the ultimate statistical mechanical origin of black hole temperature 
and entropy remains mysterious.  Here I summarize an approach in 
which the entropy is viewed as arising from ``would-be pure gauge''
degrees of freedom that become dynamical at the horizon.  For the 
(2+1)-dimensional black hole, these degrees of freedom can be counted,
and yield the correct Bekenstein-Hawking entropy; the corresponding
problem in 3+1 dimensions remains open.
\end{abstract}
\maketitle

It has been nearly 25 years since Bekenstein and Hawking first demonstrated 
that black holes are thermodynamic objects, characterized by a temperature 
and an entropy \cite{Bek,Hawking}.  Despite considerable effort, however, 
the underlying statistical mechanical source of these thermal properties 
is not yet understood.  Recent progress in string theory notwithstanding 
\cite{string}, the roots of black hole entropy remain mysterious.

Here I would like to describe a new approach to this problem, developed
over the past few years by several groups \cite{Bal,Cara,Carb,Ban}.  
This approach attributes black hole statistical mechanics to a collection 
of previously unappreciated quantum gravitational degrees of freedom,
``would-be pure gauge'' excitations that would normally be discarded as
unphysical, but that become dynamical at the black hole horizon.  The
analysis has been developed most fully for the (2+1)-dimensional black
hole of Ba{\~n}ados, Teitelboim, and Zanelli \cite{BTZ}, but there has
been a bit of progress in extending the results to other dimensions.

It is appropriate that I should present this work at a conference that
honors Tullio Regge.  Regge's work touches on this subject in two
important ways.  First, he and Claudio Teitelboim were the first to
recognize the importance of boundary terms in general relativity, and
to stress their physical significance \cite{ReggeTeit}.  Second, he and 
Jeanette Nelson were among the first to appreciate the importance of 
(2+1)-dimensional general relativity as a model for realistic quantum 
gravity, and to analyze its degrees of freedom in the Chern-Simons
formalism \cite{Nelsona,Nelsonb}.  The approach I present here can 
be viewed as a combination of these two pieces of work, albeit in a
slightly novel context.

\section{EDGE DEGREES OF FREEDOM}

We can gain some insight into the problem of black hole entropy by 
considering the simpler problem of black hole mass.  There is 
no doubt that black holes have mass.  But until 1974, it was widely 
accepted that the Hamiltonian of general relativity was simply a sum of 
constraints, and therefore vanished on physical states.  Where, then, 
could black hole mass come from?

In a seminal paper, Regge and Teitelboim resolved this problem by
showing that the Hamiltonian of general relativity must include 
boundary terms at spatial infinity \cite{ReggeTeit}.  Consider, for
instance, the role of spatial diffeomorphisms in canonical gravity.  
Let $\Sigma$ denote a constant time hypersurface, perhaps with boundary. 
An infinitesimal diffeomorphism of $\Sigma$ acts on the spatial metric 
$g_{ij}$ as 
\begin{eqnarray}
\delta g_{ij} &=& \nabla_i\xi_j + \nabla_j\xi_i \label{a1}\\
&=& \left\{ 2\int_\Sigma \nabla_l\xi_k\pi^{kl}, g_{ij} \right\} 
= \left\{ \int_\Sigma \xi_k{\cal H}^k , g_{ij} \right\} \nonumber
\end{eqnarray}
where ${\cal H}^k$ is the momentum constraint of canonical gravity,
\begin{equation}
{\cal H}^k = -2\nabla_l\pi^{kl} .
\label{a2}
\end{equation}
But the momentum constraint vanishes on physical states, so the last 
term in (\ref{a1}) is zero.  Canonical gravity thus predicts that 
physical states are diffeomorphism-invariant: the metrics $g_{ij}$
and $g_{ij}+\delta g_{ij}$ are indistinguishable.

Note, however, that the last equality of equation (\ref{a1})
involves a partial integration, which can potentially introduce a 
boundary term.  Indeed, the final Poisson bracket in (\ref{a1}) may not 
be well-defined: we should really write
\begin{equation}
\delta g_{ij} = \left\{ \left(\int_\Sigma \xi_k{\cal H}^k + 
  2\int_{\partial\Sigma} \xi_k\pi^{k\perp}\right), g_{ij} \right\} .
\label{a3}
\end{equation}
The momentum constraint has thus acquired a boundary term, which need
not vanish on physical states.  The Hamiltonian constraint is slightly
more difficult to analyze, but it, too, picks up a similar term.  Regge 
and Teitelboim showed that for a spatially open universe, the boundary 
terms in $\cal H$ and ${\cal H}^i$ give the correct ADM mass and momentum 
at infinity.

These new terms in the constraints have two implications.  First, they 
represent new observables---the ADM mass, for instance.  Second, they 
also represent new physical degrees of freedom.  In the presence of 
boundary terms, the argument for the indistinguishability of $g_{ij}$ 
and $g_{ij}+\delta g_{ij}$ no longer holds, since the right-hand side of
equation (\ref{a3}) no longer annihilates physical states.  New boundary
degrees of freedom have appeared, of the form
\begin{equation}
\delta g_{ij} = \nabla_i\xi_j + \nabla_j\xi_i , \quad 
  \left. \xi^i\right|_{\partial\Sigma}\ne 0 ,
\label{a4}
\end{equation}
which can no longer be discarded as ``pure gauge.''  The existence of 
such new degrees of freedom was already recognized by Regge and Teitelboim, 
who wrote of ``a new set of canonical pairs which describe the asymptotic 
location of the spacelike surface on which the state is defined.''

While this argument is clearest in the Hamiltonian formalism, a Lagrangian
version also exists.  A fluctuation of the spacetime metric $g_{\mu\nu}$
may be decomposed as
\begin{equation}
\delta g_{\mu\nu} = (K\xi)_{\mu\nu} + h_{\mu\nu} ,\quad 
  (K^\dagger h)_\mu = 0 
\label{a5}
\end{equation}
with
\begin{equation}
(K\xi)_{\mu\nu} = \nabla_\mu\xi_\nu + \nabla_\nu\xi_\mu ,
\label{a6}
\end{equation}
where $\nabla_\mu$ is now the full spacetime covariant derivative.  
For a closed manifold, this splitting is unique, and provides the 
standard division into ``physical'' and ``gauge'' degrees of freedom
\cite{Berger,Yorkb}.  If $M$ has a boundary, however, a unique 
decomposition requires boundary conditions that make $K^\dagger K$ 
self-adjoint.  The simplest choice is
\begin{equation}
\left. \xi^\mu\right|_{\partial\Sigma} = 0 .
\label{a7}
\end{equation}
Just as in the Hamiltonian formalism, the ``would-be gauge'' degrees of 
freedom 
\begin{equation}
\delta g_{\mu\nu} = (K\xi)_{\mu\nu} , \quad
  \left. \xi^\mu\right|_{\partial\Sigma}\ne 0 ,
\label{a8}
\end{equation}
become dynamical at the boundary. 

\section{HORIZONS AS BOUNDARIES}

The approach to black hole thermodynamics I am advocating is based on 
these same degrees of freedom, now pushed inward to the black hole 
horizon.  The obvious objection is that a horizon is not a boundary.  
This is certainly true.  Nevertheless, an event horizon in quantum 
gravity is a location at which one imposes ``boundary conditions,'' 
and these are sufficient to require the introduction of boundary terms.

Consider, for example, a question about black hole radiation.  In 
semiclassical gravity, one can ask, ``Here is a metric.  What is the 
probability of observing Hawking radiation with a given spectrum?''  In 
a full quantum theory, however, such a question makes no sense---the 
metric is a quantum variable, and cannot be fixed in advance.  Moreover, 
if one is only interested in the region near the horizon, the metric far 
from the black hole should be irrelevant.  The appropriate question is 
thus, ``Suppose the metric satisfies geometric conditions that represent
the existence of a horizon with given characteristics.  Then what is the 
probability of observing Hawking radiation with a given spectrum?''  This 
is a question about conditional probability, and the condition---the 
existence of a horizon with certain geometric properties---is a boundary 
condition.

This condition can perhaps be best understood in a path integral 
formalism.  The simplest way to impose such a requirement is to 
split the spacetime $M$ into two pieces, $M_1$ and $M_2$, along a 
hypersurface $\Sigma$, the putative event horizon. If $h$ denotes the 
metric on $\Sigma$, the total partition function is, schematically,
\begin{equation}
Z_M = \int [dh] Z_{M_1}[h]Z_{M_2}[h] ,
\label{b1}
\end{equation}
where $Z_{M_1}[h]$ and $Z_{M_2}[h]$ are the partition functions for
$M_1$ and $M_2$ with the specified induced metric $h$ on $\Sigma$, and
the integral (\ref{b1}) is restricted to boundary metrics that satisfy
the required conditions for $\Sigma$ to be a horizon.

The question is now whether the actions used to compute $Z_{M_1}[h]$
and $Z_{M_2}[h]$ should include boundary terms.  This can be answered 
by considering the  requirement of ``sewing'': if the the range of
integration in (\ref{b1}) is extended to include {\em all\/} 
intermediate metrics on $\Sigma$, the result should be equivalent 
to the ordinary path integral over $M$, independent of $\Sigma$.  This 
sewing condition has been examined for a number of exactly soluble 
systems, including free fields \cite{CCDD} and Chern-Simons theories 
\cite{Witten}, and in all cases it has been shown that the action 
must include boundary terms, guaranteeing the appearance of ``would-be 
pure gauge'' degrees of freedom at the horizon.

\section{CHERN-SIMONS THEORY}

To make this discussion less abstract, let us look at the best-understood
example, Chern-Simons gauge theory.  Let $A_\mu = A_\mu{}^aT_a$ be a gauge 
field for a nonabelian group $G$, defined on a three-manifold $M$ with 
boundary.  Fix a complex structure on the surface $\partial M$.  The 
Chern-Simons action is then
\begin{eqnarray}
I_{\hbox{\scriptsize CS}} &=& {k\over4\pi}\int_M
  \mathop{Tr}\left( A\wedge dA + {2\over3}A\wedge A\wedge A \right) 
\nonumber\\
  &+& {k\over4\pi}\int_{\partial M} \mathop{Tr}\,A_zA_{\bar z} ,
\label{c1}
\end{eqnarray}
where the boundary term is the one appropriate for fixing the component 
$A_z$ at $\partial M$. 

The equations of motion arising from this action are
\begin{equation}
 F_{\mu\nu} = 0 ,
\label{c2}
\end{equation}
where $F$ is the field strength.  On a closed topologically trivial 
manifold, Chern-Simons theory thus has no degrees of freedom.  If $M$ has 
a nontrivial fundamental group, on the other hand, the model possesses
global degrees of freedom, corresponding to Wilson loops or Aharonov-Bohm
phases around noncontractible loops.

The action (\ref{c1}) depends explicitly on the potential, and is not 
manifestly gauge invariant.  However, a simple computation shows that 
under a transformation
\begin{equation}
A = g^{-1}dg + g^{-1}\tilde A g ,
\label{c3}
\end{equation}
the action becomes 
\begin{equation}
I_{\hbox{\scriptsize CS}}[A]
   = I_{\hbox{\scriptsize CS}}[\tilde A]
   + k I^+_{\hbox{\scriptsize WZW}}[g,\tilde A_z] ,
\label{c4}
\end{equation}
where $I^+_{\hbox{\scriptsize WZW}}[g,\tilde A_z]$ is the action of
a chiral Wess-Zumino-Witten model on $\partial M$,
\begin{eqnarray}
\lefteqn{I^+_{\hbox{\scriptsize WZW}}[g,\tilde A_z]}
 \nonumber\\
 &=& {1\over4\pi}\int_{\partial M}\mathop{Tr}
 \left(g^{-1}\partial_z g\,g^{-1}\partial_{\bar z} g
 - 2g^{-1}\partial_{\bar z} g {\tilde A}_z\right) \nonumber\\
 &+& {1\over12\pi}\int_M\mathop{Tr}\left(g^{-1}dg\right)^3 .
\label{c5}
\end{eqnarray}
If $M$ is closed, the first term in (\ref{c5}) disappears, and the
second is a topological invariant, the winding number of the gauge
transformation $g: M\rightarrow G$.  For a suitably choice of $k$, 
this term always contributes an integral multiple of $2\pi$, so 
$\exp\{ iI_{\hbox{\scriptsize CS}}[A]\}$ is indeed gauge invariant.

In the presence of a boundary, however, this invariance is lost,
and the ``would-be pure gauge'' degrees of freedom become dynamical
on the boundary, with an action given by the WZW action (\ref{c5}).
These new degrees of freedom are closely related to those described 
in the first section.  Indeed, recall that the Lie derivative of 
the one-form $A_\mu dx^\mu$ satisfies the identity
\begin{equation}
{\cal L}_\xi A = d(\iota_\xi A) + \iota_\xi dA ,
\label{c6}
\end{equation}
where $\iota_\xi$ denotes the interior product.  It is then easy to
show that
\begin{equation}
{\cal L}_\xi A = D(\iota_\xi A) + \iota_\xi F ,
\label{c7}
\end{equation}
where $D$ is the gauge-covariant derivative.  On shell, the field
strength $F$ vanishes, and a diffeomorphism, the left-hand side of 
(\ref{c7}), is equivalent to a gauge transformation, the right-hand
side.  The boundary diffeomorphisms of section 1 are thus equivalent,
at least on shell, to the dynamical gauge transformations of this
section.

\section{(2+1)-DIMENSIONAL GRAVITY}

Chern-Simons theory is a fascinating model, but we are really interested
in gravity.  In three spacetime dimensions, however, we need look no
further: as Ach{\'u}carro and Townsend observed in 1986 \cite{Achu}, and 
Witten spectacularly rediscovered a few years later \cite{Wittenb}, 
(2+1)-dimensional general relativity {\em is\/} a Chern-Simons theory. 
In particular, for Lorentzian gravity with a negative cosmological 
constant $\Lambda=-1/\ell^2$, we can define an $\hbox{SU}(1,1)\times
\hbox{SU}(1,1)$ gauge field
\begin{equation}
A^\pm = \left(\omega^a \pm{1\over\ell} e^a\right) T_a ,
\label{d1}
\end{equation}
where $\omega^a = {1\over2}\epsilon^{abc}\omega_{\mu bc}dx^\mu$ is the
spin connection and $e^a = e^a{}_\mu dx^\mu$ is the triad.  The standard 
first-order form of the Einstein action can then be written as
\begin{equation}
I_{\hbox{\scriptsize grav}}
  = I_{\hbox{\scriptsize CS}}[A^+] - I_{\hbox{\scriptsize CS}}[A^-] ,
\label{d2}
\end{equation}
where $I_{\hbox{\scriptsize CS}}[A]$ is the Chern-Simons action (\ref{c1})
with a coupling constant
\begin{equation}
k = -{\ell\over4G} .
\label{d3}
\end{equation}

As in a general Chern-Simons theory, the physical degrees of freedom of
this model are Wilson loops 
\begin{equation}
R_\gamma^\pm = \mathop{Tr} \exp\int_\gamma A^\pm
\label{d3a}
\end{equation}
around closed noncontractible paths $\gamma$.  Nelson and Regge have
studied the algebra of these observables extensively \cite{Nelsona,Nelsonb},
and it is clear that they do not provide enough degrees of freedom to
account for the entropy of a (2+1)-dimensional black hole.  But by the 
discussion of the preceding section, we also expect an $\hbox{SU}(1,1)
\times\hbox{SU}(1,1)$ WZW action to be induced at the horizon of a black 
hole.  The degrees of freedom provided by this action are our candidates 
for explaining black hole statistical mechanics.

At first sight, we have been too successful: a WZW model has an infinite 
number of degrees of freedom, not the finite number needed to account
for black hole entropy.  We must be careful, however, about which states
we count as physical.  Recall that in the metric formalism, the new 
physical excitations are given by equation (\ref{a8}).  Not all boundary
diffeomorphisms appear in this equation: if $\chi$ satisfies the Killing 
equation $K\chi=0$ at $\partial M$, the right-hand side of (\ref{a8}) 
vanishes, and the corresponding constraint $\int\chi^\mu{\cal H}_\mu$
remains a genuine constraint even at $\partial M$.  In other words, a 
remnant of the Wheeler-DeWitt equation  survives at the boundary: states 
must be invariant under those diffeomorphisms that reduce to isometries at 
the horizon.

We can now proceed to count states.  I will only sketch the argument here;
the reader is referred to references \cite{Cara} and \cite{Carb} for
details.  Note first that a WZW model is a conformal field theory, and
that diffeomorphisms of $\partial M$ are therefore described by Virasoro
operators $L_n$ and ${\bar L}_n$, whose properties are well understood.  
In particular, the isometries of the horizon are rigid rotations and
time translations, which are generated by $L_0$ and $\bar L_0$, so the 
physical state condition is
\begin{equation}
L_0 |\hbox{phys}\rangle = \bar L_0 |\hbox{phys}\rangle = 0 .
\label{d4}
\end{equation}

For convenience, let us analytically continue from our $\hbox{SU}(1,1)
\times\hbox{SU}(1,1)$ WZW model with $k<0$ to the better understood
$\hbox{SL}(2,\hbox{\bf C})$ model with $k>0$.  Let $\tilde A$ denote 
the boundary values of the gauge field $A$ at the horizon, which may be 
determined from the Chern-Simons form of the classical Euclidean black 
hole solution \cite{CarTeit}.  The partition function for this model, 
\begin{equation}
Z_{\hbox{\scriptsize SL}(2,{\bf C})}(\tau)[\tilde A, \skew5\bar{\tilde A}]
  = \mathop{Tr}\left\{ e^{2\pi i\tau L_0}e^{-2\pi i\bar\tau \bar L_0}\right\} ,
\label{d5}
\end{equation}
is known from conformal field theory, and can be expressed in terms of
Weyl-Kac characters for affine $\hbox{SU}(2)$ \cite{Wittenc,Hayashi,Falc}.
Moreover, standard results from WZW theory tell us that \cite{Gepner}
\begin{equation}
Z_{\hbox{\scriptsize SL}(2,{\bf C})}(\tau)[\tilde A, \skew5\bar{\tilde A}]
  = \sum \rho(N,\bar N)q_1^{N-\bar N}q_2^{N+\bar N} ,
\label{d6}
\end{equation}
where $q_1=e^{2\pi i\tau_1}$, $q_2=e^{-2\pi\tau_2}$, and $\rho(N,\bar N)$
is the number of states for which the Virasoro generators $L_0$ and $\bar
L_0$ have eigenvalues $N$ and $\bar N$.  The number of states satisfying
the physical state condition (\ref{d4}) is thus $\rho(0,0)$, which can
be extracted from (\ref{d6}) by contour integration.

This computation is carried out in reference \cite{Carb}.  The outcome
is that
\begin{equation}
\ln\rho(0,0) = {2\pi r_+\over4G} + {\pi r_+\over\ell} + \dots ,
\label{d7}
\end{equation}
where $r_+$ is the radius of the event horizon.  The first term in this
expression is precisely the right Bekenstein-Hawking entropy, while the
second is a one-loop correction.  Our counting argument has thus succeeded.
A similar computation can be performed directly in Lorentzian signature,
again yielding the correct entropy \cite{Cara}.

\section{THE REAL WORLD}

The evidence from 2+1 dimensions is certainly suggestive, but it is 
not conclusive.  Te obvious question is whether these results can be
generalized to 3+1 dimensions.  In this simple form, they certainly 
cannot.  The Chern-Simons formulation of (2+1)-dimensional gravity
allowed us to trade the complicated diffeomorphism group for a much
simpler gauge group, via equation (\ref{c7}).  No such procedure is
known in 3+1 dimensions, and we have no simple splitting of the action 
into ``bulk'' and ``boundary'' terms comparable to that of equation 
(\ref{c4}).  On the other hand, the arguments of section 1 hold in any 
number of dimensions.  There are certainly ``would-be pure gauge'' 
degrees of freedom in 3+1 dimensions; the problem is that we do not 
know how to count them.

One interesting place to test these ideas is (1+1)-dimensional dilaton 
gravity.  It is not too hard to show that a suitable choice of boundary 
conditions induces a dynamical theory on the horizon of a (1+1)-dimensional 
black hole, but no analog of the physical state conditions (\ref{d4}) is 
yet known.  It is also possible that state-counting arguments in the 
(3+1)-dimensional loop representation are looking at these ``would-be 
pure gauge'' degrees of freedom \cite{Rovelli}, but the connection is 
still somewhat speculative.

\end{document}